\documentclass[%
 reprint,
 superscriptaddress,
 amsmath,amssymb,
 aps,
 pra,
]{revtex4-2}
\usepackage{hyperref}
\usepackage{amsmath}
\usepackage{xcolor}
\usepackage{graphicx}
\usepackage{siunitx}
\usepackage{bm}
\usepackage{mathtools}
\usepackage{upgreek}

\DeclarePairedDelimiter\bra{\langle}{\rvert}
\DeclarePairedDelimiter\ket{\lvert}{\rangle}
\DeclarePairedDelimiterX\braket[2]{\langle}{\rangle}{#1\,\delimsize\vert\,\mathopen{}#2}

\usepackage{datetime}

\newcommand{\bsigma}{\bm{\sigma}}

\date{\today;\currenttime}
\begin{document}

\title{A Cluster-Based Model of the Spectrum of Erbium-Doped GdVO\(_4\)}
\author{Zachary H. Roberts}
\author{Masaya Hiraishi}
\altaffiliation{Now at: Basic Research Laboratories, NTT, Inc., 3-1 Morinosato Wakamiya, Atsugi-shi, Kanagawa 243-0198, Japan}
\author{Luke S. Trainor}
\author{Jevon J. Longdell}
\email{jevon.longdell@otago.ac.nz}
\affiliation{%
  Department of Physics, University of Otago, Dunedin, New Zealand}
\affiliation{Dodd-Walls Centre for Photonic and Quantum Technologies, Dunedin, New Zealand}

\begin{abstract}
Experimental observations of rare-earth ions doped into an antiferromagnetic crystal show an enriched optical spectrum. In this paper we present a cluster-based model to describe erbium ions doped into a gadolinium vanadate (Er:GdVO\(_4\)) host crystal, wherein the erbium ion couples directly to its four nearest neighbour gadolinium ions, which in turn couple to the mean field of the rest of the crystal. Compared to previous models in the literature, the parameters used to fit this model are fewer in number, with clearer physical origins. Agreement with the experimentally observed optical spectrum of Er:GdVO\(_4\) suggests that our model succeeds in capturing the most important interactions of the system, suggesting that it may be useful for predicting microwave-to-optical transduction in future experiments.
\end{abstract}

\maketitle

\section{Introduction}

Rare-earth dopants in solid-state crystals provide a promising platform for realising both long optical coherence times \cite{Rancic_2018, Bottger_2009} and microwave to optical transduction \cite{Awschalom_2021, Everts_2019, OBrien_2014, Fernandez-Gonzalvo_2019, Xie_2025}. The \(4f\) electrons that are characteristic of rare-earth ions are enclosed within outer shells of the ion, electrostatically shielding them from the crystal environment. This gives \(4f\)-to-\(4f\) transitions very narrow linewidths, and also means that at low temperatures the dominant dephasing mechanism for both optical and spin transitions is magnetic fluctuations from spins in the host crystal \cite{Zhong_2015, Rancic_2018}.

In recent work performed on erbium-doped gadolinium vanadate (Er:GdVO\(_4\)) we showed that by doping the rare-earth ion into a host crystal that magnetically orders below a critical temperature provides a quiet environment in which electronic spin flips in the host crystal are suppressed, leading to long optical coherence times \cite{Hiraishi_2025}. A further property of magnetically ordered systems is that they display magnetic resonance in the form of magnon modes. Measurements of the \(\bm{k}=\bm{0}\) magnons in GdVO\(_4\) suggest that at zero applied field,  the magnon frequency will occur at 30.61 GHz \cite{Abraham_1992}, well within the microwave regime, and thus coupling between magnons and doped rare-earth ions in magnetically ordered crystals presents an intriguing mechanism for realising microwave to optical transduction \cite{Lambert_2020}. Supporting this idea, in our previous paper, we observed avoided crossings in the optical spectra which we were able to attribute using a simple model to strong coupling between the erbium ions and gadolinium spins in the host crystal.

Previous interest in the effect of an antiferromagnetic host on rare earth dopants has been in the context of electron spin resonance \cite{Bleaney_1991, Abraham_1992_Dy, Bleaney_1982}. In these papers, the authors model the electron as being exposed to an effective field that is the sum of both external and internal (dipole and exchange) fields present at the dopant site. On one sublattice, the two field components will be parallel, while on the other they will be antiparallel, thus explaining the splitting of the resonances at nonzero applied field. A similar idea is used in Ref. \cite{Berrington_Thesis}, wherein the optical spectrum of erbium in ErLiF\(_4\) is modelled by starting with a crystal field theory, and then introducing a Zeeman term describing the interaction between the erbium ion and the effective field. As discussed in Ref. \cite{Hiraishi_2025}, such models do a good job of explaining the energy levels of the rare-earth ion by itself, but they fail to match the observed optical spectra as they do not include excitations of the gadolinium spin lattice, which couple to the erbium levels at higher frequencies.

In our previous work \cite{Hiraishi_2025}, we modelled the Er:GdVO\(_4\) system by placing the erbium ion in a crystal field environment, as well as a static mean field consisting of dipole and exchange fields from the gadolinium lattice. We then incorporated the gadolinium excitations `by hand' by coupling the erbium to a harmonic oscillator. In this paper, we build this previous model by allowing the erbium to couple directly to its four nearest neighbour gadolinium ions, which in turn couple to the dynamic mean field of the bulk crystal. This was inspired by cluster model approaches to explain the spin wave spectrum of doped magnetic insulators \cite{Cowley_1972}. Not only does this offer an improved fit to experimental data, but it allows us to extend the model to consider cases where an external field is applied off-axis, and to higher applied fields where phase transitions in the magnetic ordering are observed.

This paper is organised as follows. In Section \ref{sec:Methods} we discuss the crystallographic and magnetic properties of gadolinium vanadate, before deriving the Hamiltonian we use to describe our coupled erbium-crystal system. We also discuss the method we use to find the mean field spin configuration of the bulk crystal. In Section \ref{sec:Results} we present the output of our `cluster model' alongside our measured transmission spectrum for Er:GdVO\(_4\). Finally, in Section \ref{sec:Discussion} we discuss the successes of the model, as well as shortcomings and potential further refinements that could be made.

\section{Methods}\label{sec:Methods}

\subsection{Er:GdVO\(_4\)}

\begin{figure*}
    \centering
    \includegraphics[scale=0.7]{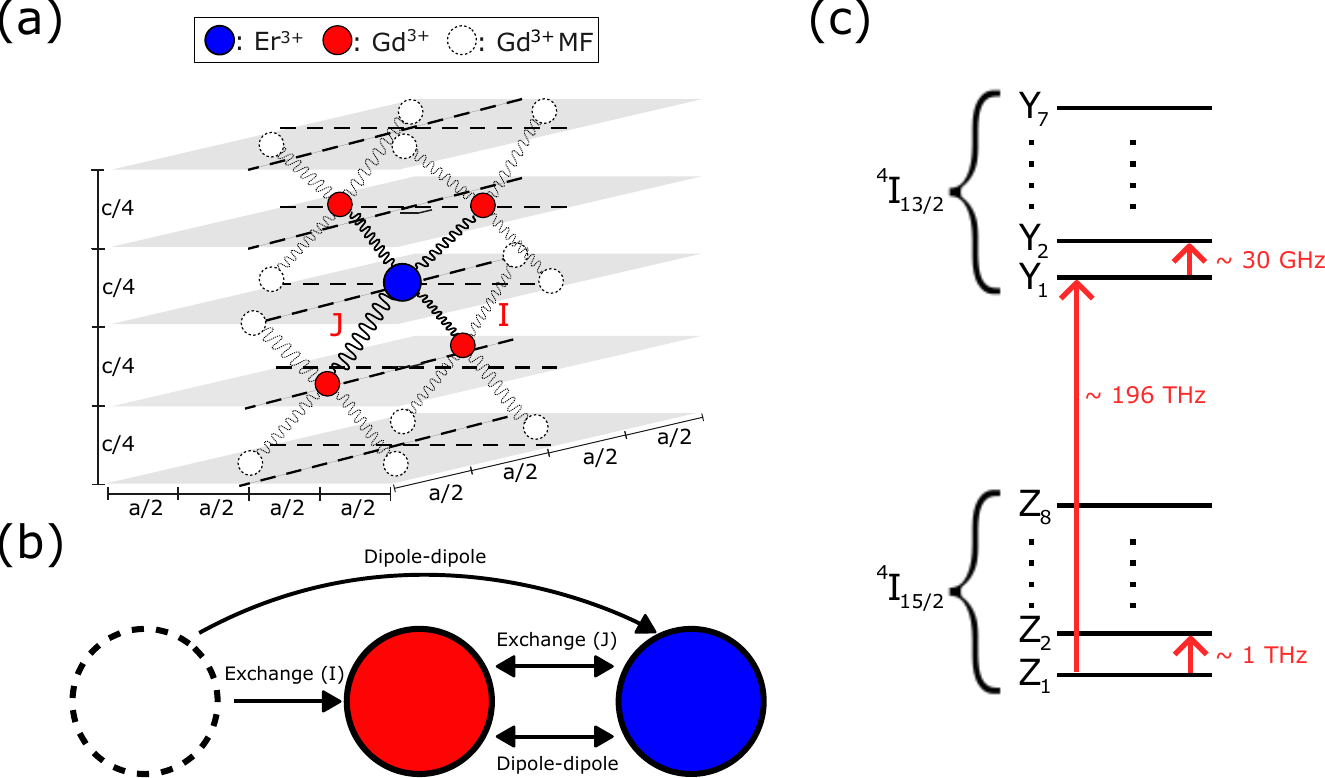}
    \caption{(a) Local environment of the erbium dopants, which occupy a site with \(\text{D}_{2d}\) symmetry. Here we have labelled erbium ion (blue), its four nearest neighbour gadolinium ions (red), and the surrounding mean field (MF) gadolinium ions (white). Also shown are the Er-Gd exchange interaction (J) and Gd-Gd exchange interaction (I). Figure adapted from Figure 1(a) in Ref. \cite{Hiraishi_2025}. (b) Schematic showing the interactions between the erbium ion, nearest neighbour gadolinium ions, and mean field gadolinium ions. (c) Energy level diagram of the \(^4I_{15/2}\) and \(^4I_{13/2}\) levels of the erbium ion, split into Kramers doublets by the crystal field. The transition frequencies are from Ref. \cite{Bertini_2004}. The splitting between \(Z_1\) and \(Z_2\) is large enough that we can ignore \(Z_2\), but \(Y_1\) and \(Y_2\) are close enough in energy to be mixed. Interactions with the gadolinium spins serves to remove the degeneracy of the doublets, giving rise to four transitions at zero applied magnetic field: \(2\times (Z_1\rightarrow Y_1)\) and \(2\times (Z_1\rightarrow Y_2)\).}
    \label{fig:crystal}
\end{figure*}

The material we focus on in this work is erbium-doped gadolinium orthovanadate (Er:GdVO\(_4\)). As shown in Figure \ref{fig:crystal}(a), the unit cell for this crystal is tetragonal, with dimensions \(a = 7.2126\) Å and \(c = 6.3485\) Å \cite{Abraham_1992}. Each gadolinium ion occupies a site with \(\text{D}_{2d}\) symmetry \cite{Milligan_1952}. The crystal orders antiferromagnetically below a a Néel temperature of \(T_N=2.495\) K \cite{Cashion_1970}. The magnetic structure is composed of two sublattices that align parallel and antiparallel to the easy axis (\(c\) axis). The structure is bipartite: each gadolinium ion on one sublattice is surrounded by four nearest neighbours belonging to the other sublattice.

\subsection{Model}

We build upon the model developed in Ref. \cite{Hiraishi_2025} by considering an erbium ion directly interacting with its four nearest neighbour gadolinium ions, forming a cluster of five atoms as shown in Figure \ref{fig:crystal}(a). This cluster then interacts with the mean field created by the rest of the lattice. The Hamiltonian for this system is:
\begin{equation}\label{eq:HSys}
    H = H_\text{Er}+H_\text{Gd}+H_\text{Er-Gd}
\end{equation}
Where \(H_\text{Er}\) is the erbium Hamiltonian, consisting of free ion, crystal field, and Zeeman terms:
\begin{equation}\label{eq:ErH}
    H_\text{Er}=H_\text{FI, Er}+H_\text{CF, Er}+H_\text{Z, Er}
\end{equation}
The Zeeman Hamiltonian takes the form:
\begin{equation}
    H_\text{Z, Er} = \mu_B\bm{B}\cdot(\bm{L}_\text{Er}+2\bm{S}_\mathrm{Er})
\end{equation}
Here \(\mu_B\) is the Bohr magneton, \(\bm{B}\) is the applied magnetic field, and \(\bm{L}_\text{Er}\) and \(\bm{S}_\text{Er}\) are the orbital and spin angular momentum operators for the erbium ion.

We write the Hamiltonian for the lattice of gadolinium ions as:
\begin{align}\label{eq:GdHam}
    H_\text{Gd}  = & \ h\gamma\sum_{i,j}\bm{B}\cdot\bm{S}_{i,j}-I\sum_{\langle i,j\rangle}\bm{S}_i\cdot\bm{S}_j -D\sum_{i,j}(S^{z}_{i,j})^2
\end{align}
The first term is the Zeeman interaction, with \(\gamma=g\mu_B/h\) being the gyromagnetic ratio, \(g=1.991\) (isotropic) being the g-factor for gadolinium \cite{Abraham_1992}, \(\mu_B\) being the Bohr magneton, and \(h\) being Planck's constant. The quantity \(\bm{S}_{i,j}\) is the spin operator for the gadolinium ion at site \(i/j\). As gadolinium vanadate arranges antiferromagnetically in the ground state, we use the \(i\) indices to label the up sublattice and the \(j\) indices to label the down sublattice. The second term is a Heisenberg exchange interaction, with \(I\) being the coupling energy (negative for an antiferromagnetic crystal), and \(\langle i,j\rangle\) denoting nearest neighbours. The third term is the anisotropy term which causes the spins to align along the \(z\)-direction, with \(D\) parameterising the strength of this effect. The value of \(D\) can be obtained from the known anisotropy field \(H_A=0.401\) T \cite{Abraham_1992} using the relation \cite{Rezende}:
\begin{equation}
    D=\frac{\gamma h H_\text{A}}{2S}
\end{equation}
Where \(S=7/2\) is the magnitude of the gadolinium spins. Abraham \textit{et al.} (1992) \cite{Abraham_1992} also give a value for the exchange field \(H_E=1.655 T\), which is related to the exchange coupling (\(I\)) by a similar relation. However, in this work we include \(I\) as a fitting parameter, and use it to calculate the exchange field as \cite{Rezende}:
\begin{equation}
    H_\text{E} = \frac{2Sz|I|}{\gamma h}
\end{equation}
Just for this equation, \(z\) is the number of nearest neighbours.

Finally, the Hamiltonian for the interaction between the erbium ion and the gadolinium lattice can be written:
\begin{align}
\begin{split}
    H_\text{Er-Gd}  = & \ -2J\sum_\delta\bm{S}_\text{Er}\cdot\bm{S}_\delta \\ &-\sum^N_{i\neq0}\frac{\mu_0}{4\pi r_{i0}^3}\left(3(\boldsymbol{\mu_i\cdot\hat{\bm{r}}_{i0}})(\boldsymbol{\mu}_\text{Er}\cdot\hat{\bm{r}}_{i0})-\boldsymbol{\mu}_i\cdot\boldsymbol{\mu}_\text{Er} \right) 
\end{split} \\
\begin{split}
     = & \ -2J\sum_\delta\bm{S}_\text{Er}\cdot\bm{S}_\delta \\ &-g\mu_B^2\sum^N_{i\neq0}\frac{\mu_0}{4\pi r_{i0}^3}\bigg(3(\boldsymbol{\bm{S}_i\cdot\hat{\bm{r}}_{i0}})\left((\boldsymbol{L}_\text{Er}+2\bm{S}_\text{Er})\cdot\hat{\bm{r}}_{i0}\right) \\ & -\bm{S}_i\cdot(\boldsymbol{L}_\text{Er}+2\boldsymbol{S}_\text{Er}) \bigg)
\end{split}
\end{align}
The first term represents a Heisenberg exchange interaction with coupling \(J\) between the erbium ion, located at site \(i=0\), and its nearest neighbours, labelled by \(\delta\). The second term represents the dipole-dipole interaction between the erbium and all the gadolinium ions in the lattice, with \(\bm{r}_{i0}\) being the separation vector between the erbium ion and the \(i\)th gadolinium ion,  and \(\boldsymbol{\mu}\) denoting magnetic dipole vectors, which for gadolinium is just \(-g\mu_B\bm{S}\) since this ion has \(\bm{L}=\bm0\).

In the ground state, the gadolinium lattice arranges antiferromagnetically, giving rise to two sublattices. Excitations to this ground state thus take the form of `spin deviations', which we represent with creation (annihilation) operators \(a_i^\dagger\) (\(a_i\)) and \(b_j^\dagger\) (\(b_j\)) for the two sublattices. For arbitrary magnetic field direction, these two sublattices will in general point along different quantisation axes which we denote \(\hat{\bm{z}}'\) and \(\hat{\bm{z}}''\). To complete the coordinate systems, we choose \(\hat{\bm{y}}'=\hat{\bm{y}}''=(\hat{\bm{z}}'\times\hat{\bm{z}}'')/|\hat{\bm{z}}'\times\hat{\bm{z}}''|\), and then \(\hat{\bm{x}}'=\hat{\bm{y}}'\times\hat{\bm{z}}'\) and \(\hat{\bm{x}}''=\hat{\bm{y}}''\times\hat{\bm{z}}''\).

Keeping terms no greater than quadratic in order, we write our gadolinium spin operators in terms of these spin deviation operators as (See Rezende Chapter 5.3 \cite{Rezende}):
\begin{align}
    S_i^+=\sqrt{2S}a_i && S_i^-=\sqrt{2S}a_i^\dagger && S_i^{z'}=S-a_i^\dagger a_i \label{eq:UpDev}\\
    S_j^+=\sqrt{2S}b_j && S_j^-=\sqrt{2S}b_j^\dagger && S_j^{z''}=S-b_j^\dagger b_j \label{eq:DownDev}
\end{align}
Before substituting these into Equation~\eqref{eq:GdHam} we make the cluster approximation: namely, we only treat the four nearest neighbours as individual gadolinium ions, which then interact with the mean spin field of the rest of the gadolinium crystal. These nearest neighbours will all be on the same sublattice, but they may be on either the up or down sublattice. Equation~\eqref{eq:GdHam} then becomes:
\begin{equation}
\begin{split}
    H_\text{Gd} = & \ h\gamma\sum_{i,j}\bm{B}\cdot\bm{S}_{i,j}   -4I\sum_{j}\langle\bm{S}_1\rangle\cdot\bm{S}_j - 4I\sum_{i}\bm{S}_i\cdot\langle\bm{S}_2\rangle \\ & - D\sum_{i,j}(S^{z}_{i,j})^2
\end{split}
\end{equation}
Letting 1 and 2 denote the up and down sublattices respectively. We can now split up the full Hamiltonians into individual Hamiltonians for each of the nearest neighbour gadolinium ions on the two spin sublattices:
\begin{equation}\label{eq:HGd1}
\begin{split}
    H_\text{Gd}^{i/j} = & \ \frac{1}{2}\left[(h\gamma\bm{B}-6IS\bsigma_{2/1})\cdot\hat{\bm{x}}'^/{''}\right]\sqrt{2S}\left(a_{i/j}+a_{i/j}^\dagger \right)\\ & - \frac{i}{2}\left[(h\gamma\bm{B}-6IS\bsigma_{2/1})\cdot\hat{\bm{y}}'^/{''}\right]\sqrt{2S}\left(a_{i/j}-a_{i/j}^\dagger \right) \\ &  - \left[(h\gamma\bm{B}-6IS\bsigma_{2/1})\cdot\hat{\bm{z}}'^/{''}\right]a_{i/j}^\dagger a_{i/j}  -D\left(S^{z}_{i/j}\right)^2
\end{split}
\end{equation}
Here we have dropped constant terms and terms that are higher than quadratic order in the spin deviation operators. We have also made the substitution \(\langle\bm{S}_{1/2}\rangle=S\bsigma_{1/2}\), where \(S=7/2\) is the magnitude of the gadolinium spins, and \(\bsigma_0=[\bsigma_1 \  \bsigma_2]\) is the 6-dimensional (3 dimensions for each sublattice) equilibrium spin configuration for the mean field spins, which we calculate in Section \ref{sec:Minimisation}. To get the factor of \(6I\) we first replace \(4I\rightarrow 3I\) as each of the nearest neighbour gadolinium ions only sees 3 other gadolinium ions. We then introduce a factor of 2 to ensure we still double count each Gd-Gd interaction.

For both sublattices, the lowest two energy levels for the individual gadolinium ions will be \(\ket{S_{z'/z''}}=\ket{\frac{7}{2}}\) and \(\ket{S_{z'/z''}}=\ket{\frac{5}{2}}\). It is easy to show that in a basis formed by these two states, our spin deviation operators take the form:
\begin{align}
    a_i=\begin{bmatrix}
        0 & 1 \\
        0 & 0
    \end{bmatrix}_{i} && b_j=\begin{bmatrix}
        0 & 1 \\
        0 & 0
    \end{bmatrix}_{j}
\end{align}
Where the subscript denotes which ion subspace the matrix acts on.
\begin{figure*}
    \centering
    \includegraphics[scale=0.5]{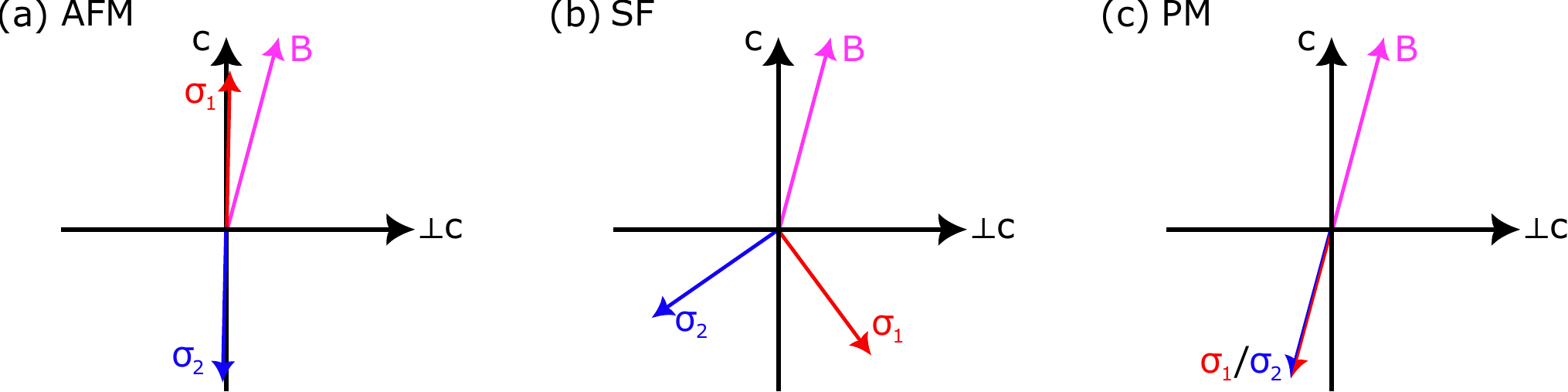}
    \caption{The three different spin configuration phases our antiferromagnetic system can exhibit. The black arrows depict the crystal axes, \(\bm{B}\) represents the external magnetic field, and \(\bsigma_1\) and \(\bsigma_2\) represent the spin orientations of the two spin sublattices. (a) Antiferromagnetic phase (AFM). (b) Spin-flop phase (SF), \(B>\left(2H_EH_A+H_A^2\right)^{1/2}\) \cite{Rezende}. (c) Paramagnetic phase (PM), \(B>2H_E-H_A\).}
    \label{fig:SpinPhas}
\end{figure*}
Assuming the erbium only experiences exchange interactions with its nearest neighbours, the erbium ion does not exchange with the mean field of gadolinium spins. It does still interact with them through the longer-range dipole-dipole interaction. This leads us to modify Equation~\eqref{eq:ErH} as:
\begin{equation}\label{eq:HErFull}
H_\text{Er}=H_\text{FI, Er}+H_\text{CF, Er}+H_\text{Z, Er}+H_\text{dip}
\end{equation}
Where:
\begin{equation}
\begin{split}
    H_\text{dip} = & -g\mu_B^2\sum^N_{i\notin \{0,\delta\}}\frac{\mu_0}{4\pi r_{i0}^3}\Big(3(\bsigma_i\cdot\hat{\bm{r}}_{i0})\left((\boldsymbol{L}_\text{Er}+2\bm{S}_\text{Er})\cdot\hat{\bm{r}}_{i0}\right) \\ & -\bsigma_i\cdot(\boldsymbol{L}_\text{Er}+2\boldsymbol{S}_\text{Er}) \Big),
\end{split}
\end{equation}
accounting for the dipole--dipole interactions between the erbium ion and the other ions in the crystal, but excluding those from the cluster.
For the interaction Hamiltonian, we are just left with the terms in the cluster:
\begin{widetext}
\begin{equation}\label{eq:HCoupling}
\begin{split}
    H_\text{Er-Gd}= & \ -2J\sum_\delta\bm{S}_\text{Er}\cdot\bm{S}_\delta-g\mu_B^2\sum_{\delta}\frac{\mu_0}{4\pi r_{\delta0}^3}\bigg(3(\bm{S}_\delta\cdot\hat{\bm{r}}_{\delta0})\left((\boldsymbol{L}_\text{Er}+2\bm{S}_\text{Er})\cdot\hat{\bm{r}}_{\delta0}\right) -\bm{S}_\delta\cdot(\boldsymbol{L}_\text{Er}+2\boldsymbol{S}_\text{Er}) \bigg) \\
    = & \ -2J\sum_\delta\bm{S}_\text{Er}\cdot\left(\frac{1}{2}\sqrt{2S}(a_\delta+a_\delta^\dagger)\hat{\bm{x}}^{\alpha}-\frac{i}{2}\sqrt{2S}(a_\delta-a_\delta^\dagger)\hat{\bm{y}}^{\alpha}+(S-a_\delta^\dagger a_\delta)\hat{\bm{z}}^{\alpha}\right) \\ & \ -g\mu_B^2\sum_\delta \frac{\mu_0}{4\pi r_{\delta0}^3} \Bigg(3\left(\left(\frac{1}{2}\sqrt{2S}(a_\delta+a_\delta^\dagger)\hat{\bm{x}}^{\alpha}-\frac{i}{2}\sqrt{2S}(a_\delta-a_\delta^\dagger)\hat{\bm{y}}^{\alpha}+(S-a_\delta^\dagger a_\delta)\hat{\bm{z}}^{\alpha}\right)\cdot\hat{\bm{r}}_{\delta0}\right)\hat{\bm{r}}_{\delta0}\cdot(\bm{L}_\text{Er}+2\bm{S}_\text{Er}) \\ & \ - \left(\frac{1}{2}\sqrt{2S}(a_\delta+a_\delta^\dagger)\hat{\bm{x}}^{\alpha}-\frac{i}{2}\sqrt{2S}(a_\delta-a_\delta^\dagger)\hat{\bm{y}}^{\alpha}+(S-a_\delta^\dagger a_\delta)\hat{\bm{z}}^{\alpha}\right) \cdot \left(\bm{L}_\text{Er}+2\bm{S}_\text{Er}\right) \Bigg)
\end{split}
\end{equation}
\end{widetext}
Where \(x^\alpha=x'\) or \(x''\). The interactions between the erbium, gadolinium, and mean field gadolinium ions are summarised in Figure \ref{fig:crystal}(b).

\subsection{Bulk Spin Configuration}\label{sec:Minimisation}

To find the equilibrium spin configuration of the bulk GdVO\(_4\) crystal, we treat the gadolinium spin lattice as consisting of two sublattices with spins \(\bm{S}_1\) and \(\bm{S}_2\). Furthermore, we treat the spins classically, and rewrite them as \(\bm{S}_i=S\bsigma_i\), where \(S=7/2\) is the magnitude of the gadolinium spins, and \(\bsigma_i\) is a unit vector giving the spin direction. Using Equation~\eqref{eq:GdHam} we can write the energy for a given spin configuration as:
\begin{equation}\label{eq:ClassE}
\begin{split}
    \frac{E}{h\gamma S} =& \ 2\mu_BH_E\bsigma_1\cdot\bsigma_2-\mu_BH_A\left(\left(\sigma_1^z\right)^2+\left(\sigma_2^z\right)^2\right) \\ &+ \mu_B g \bm{B}\cdot\left(\bsigma_1+\bsigma_2\right)
\end{split}
\end{equation}
We can then find the equilibrium spin configuration \(\bsigma_0=[\bsigma_1 \ \bsigma_2]\) by finding the components of \(\bsigma_0\) that minimise Equation~\eqref{eq:ClassE}, subject to the normalisation constraint \(\bsigma_1\cdot\bsigma_1=\bsigma_2\cdot\bsigma_2=1\). In practice, this is done using a basinhopping algorithm with gradient descent occurring after each hop.

In Figure \ref{fig:SpinPhas} we show the three magnetic phases that the system may exhibit: antiferromagnet (Figure \ref{fig:SpinPhas}(a)), spin-flop (Figure \ref{fig:SpinPhas}(b)), and paramagnet (Figure \ref{fig:SpinPhas}(c)).

\subsection{Fitting}

\begin{figure*}[t]
    \centering
    \includegraphics[scale=0.5]{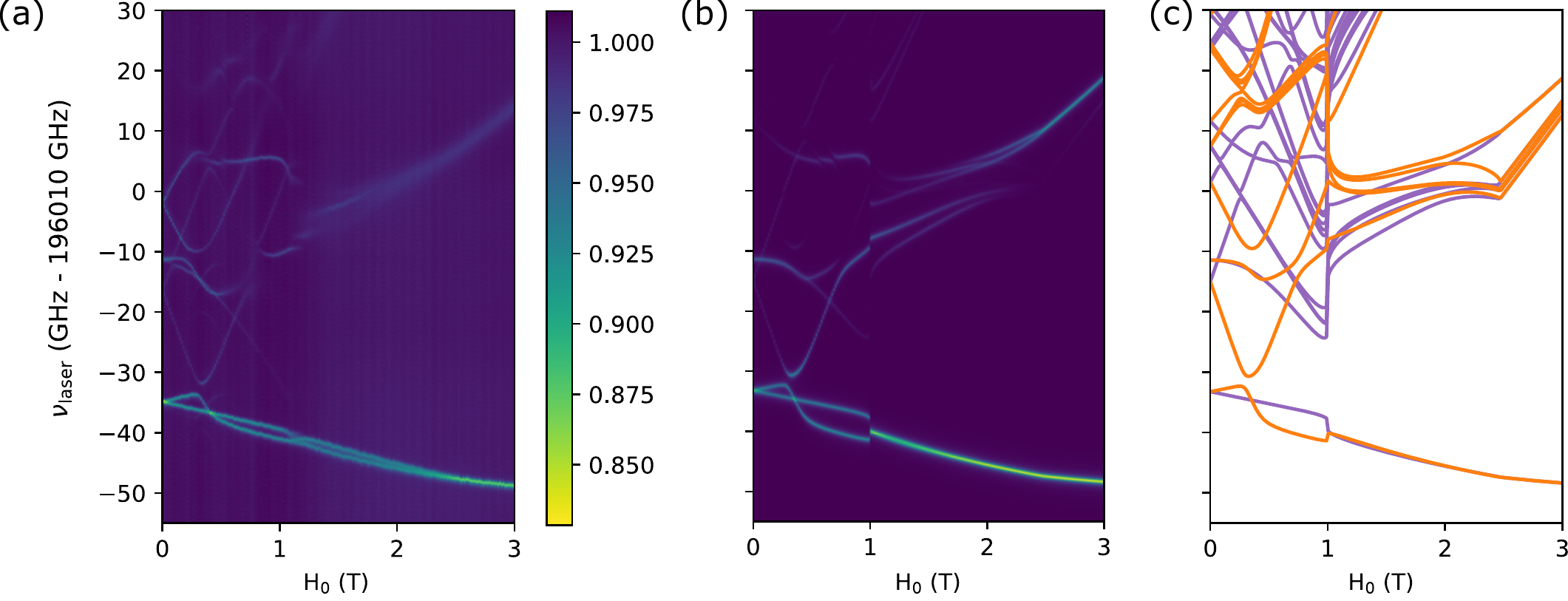}
    \caption{(a) Optical transmission spectrum with light polarised halfway between the \(a\) and \(c\)-axes of the crystal, and applied magnetic field misaligned by \qty{6.8}{\degree} from the \(c\)-axis. (b) Output of our cluster model, with relative transition strengths estimated using equation \ref{eq:strengths}. (c) Model output without transition strengths being taken into account. Orange lines indicate that the erbium is surrounded by gadolinium ions on the `down' sublattice, while purple lines indicate that the erbium is surrounded by gadolinium ions on the `up' sublattice. Here `up' and `down' point along the crystallographic \(c\)-axis.}
    \label{fig:On-axis}
\end{figure*}

The full list of parameters we fit are shown in Table \ref{tab: Cluster}. This list includes the crystal-field parameters $B^{(k)}_q$ relevant to $\text{D}_{2d}$ symmetry, Slater parameters $F^k$, spin-orbit coupling $\zeta$, as well as the electronic exchange parameters \(I\) and \(J\). We also include \(E_\text{corr}\), an offset of the \(^4I_{13/2}\) levels relative to the \(^4I_{15/2}\) levels that we introduce to get the model to agree better with the data without changing the coarse structure of the energy levels, which agrees well with values from \textcite{Bertini_2004}. We use the previously found misalignment angle of \(\theta = \qty{6.8}{\degree}\). The remaining free-ion parameters, which we do not vary, are those for LaF\(_3\) found by \textcite{Carnall_1989}. These values are not expected to vary much between hosts. To fit these parameters we generate transition frequencies for the \(Z_1\rightarrow Y_1/Y_2\) transitions, and compare them to the measured spectrum shown in Figure \ref{fig:On-axis}(a). As starting parameters we use the parameters we found in our previous paper \cite{Hiraishi_2025}, and then use a basinhopping algorithm to minimise the difference between the calculated and experimental values.

Using the data shown in Figure \ref{fig:On-axis}(a), we find the optimum values for our fit parameters. 
\begin{table}[tb]
\caption{Fitted values of parameters from the cluster model.}
    \centering
    \begin{tabular}{c r l}
        Parameter & value\hspace{1ex} & \,unit  \\
        \hline\hline
        $E_0$ & $35539.7$ & \unit{\per\centi\meter}  \\
        $F^2$ & $97123.5$ & \unit{\per\centi\meter}  \\
        $F^4$ & $66243.2$ & \unit{\per\centi\meter}  \\
        $F^6$ & $54049.1$ & \unit{\per\centi\meter}  \\
        $\zeta$ & $2362.7$ & \unit{\per\centi\meter}  \\
        $B_2^0$ & $-94.2$ & \unit{\per\centi\meter}  \\
        $B_4^0$ & $317.8$ & \unit{\per\centi\meter}  \\
        $B_6^0$ & $-610.7$ & \unit{\per\centi\meter}  \\
        $B_4^4$ & $769.4$ & \unit{\per\centi\meter}  \\
        $B_6^4$ & $-45.3$ & \unit{\per\centi\meter}  \\
        $E_\mathrm{corr.}$ & $-6.4$ & \unit{\per\centi\meter}  \\
        $I$ & $-0.0475$ & \unit{\per\centi\meter}  \\
        $J$ & $-0.0252$ & \unit{\per\centi\meter}  \\
    \end{tabular}
    \label{tab: Cluster}
\end{table}

\subsection{Transition Strengths}

The \(4f\)-to-\(4f\) electronic transitions of rare-earth ions should in principle be electric-dipole forbidden, as the wavefunction parity is not changed, but may be magnetic-dipole allowed \cite{Wybourne_and_Smentek}.
In practice, the states are mixed with other states of opposite parity, and electric-dipole transitions become allowed again.
These \emph{forced} electric-dipole transitions are often of a similar order of magnitude to allowed magnetic-dipole transitions, but they are highly environment sensitive and difficult to predict \cite{Duan_2006}.
For $^4I_{15/2}\rightarrow{}^4I_{13/2}$ transitions of erbium ions oscillator strengths for electric- and magnetic- dipole transitions are typically of the order $f_\text{ED}\sim\num{e-6}$, $f_\text{MD}\sim\num{e-7}$ \cite{berringtonNegativeRefractiveIndex2023}.
From our crystal-field model, we can calculate the transition strengths of the magnetic-dipole components of the transitions by calculating the off-diagonal matrix elements of the magnetic moment:
\begin{equation}\label{eq:strengths}
    P \propto|\bra{f}\bm{L}_\text{Er}+2\bm{S}_\text{Er}+\sum_\delta g\bm{S}_\delta\ket{i}|^2,
\end{equation}
where \(P\) denotes the transition strength, and \(\ket{f}\), \(\ket{i}\) denote the final and initial states of the system, respectively.
The electric-dipole contributions are not calculable from our model, however their selection rules can still be derived \cite{marinoEnergyLevelStructure2016}.

\section{Results}\label{sec:Results}

In Figure \ref{fig:On-axis} we compare the optical transmission spectrum of Er:GdVO\(_4\) we performed previously \cite{Hiraishi_2025} with the output of the cluster model, with field misalignment set to \(\theta=\qty{6.8}{\degree}\), where $\theta$ is the angle between the applied field and the $c$ axis. In Figure \ref{fig:On-axis}(c) we show just the transition frequencies calculated by the model, without taking transition strengths into account. We see that the cluster model is able to accurately replicate the transition spectrum of the material at low fields when the crystal is in the antiferromagnetic phase, as well as at higher fields during the spin flop and paramagnetic phases of the material. Our method for estimating transition strengths matches the transition strengths observed in the data reasonably well, with the notable exception that the fourth energy level is predicted to be much weaker than what we observe. However, as shown in Figure \ref{fig:On-axis}(c), this transition is still present in the model.

At field values of just past \(1\) T, the measured transmission spectrum undergoes an abrupt change, with most of the spectral lines vanishing as the crystal enters the spin flop phase. Our model is able to replicate this behaviour, and we see that a lot of the spectral lines rapidly converge past \(1\) T. This leads to stronger absorption of the few lines that remain, which matches what we see in the measured spectrum.

\begin{figure*}[t]
    \centering
    \includegraphics[scale=0.5]{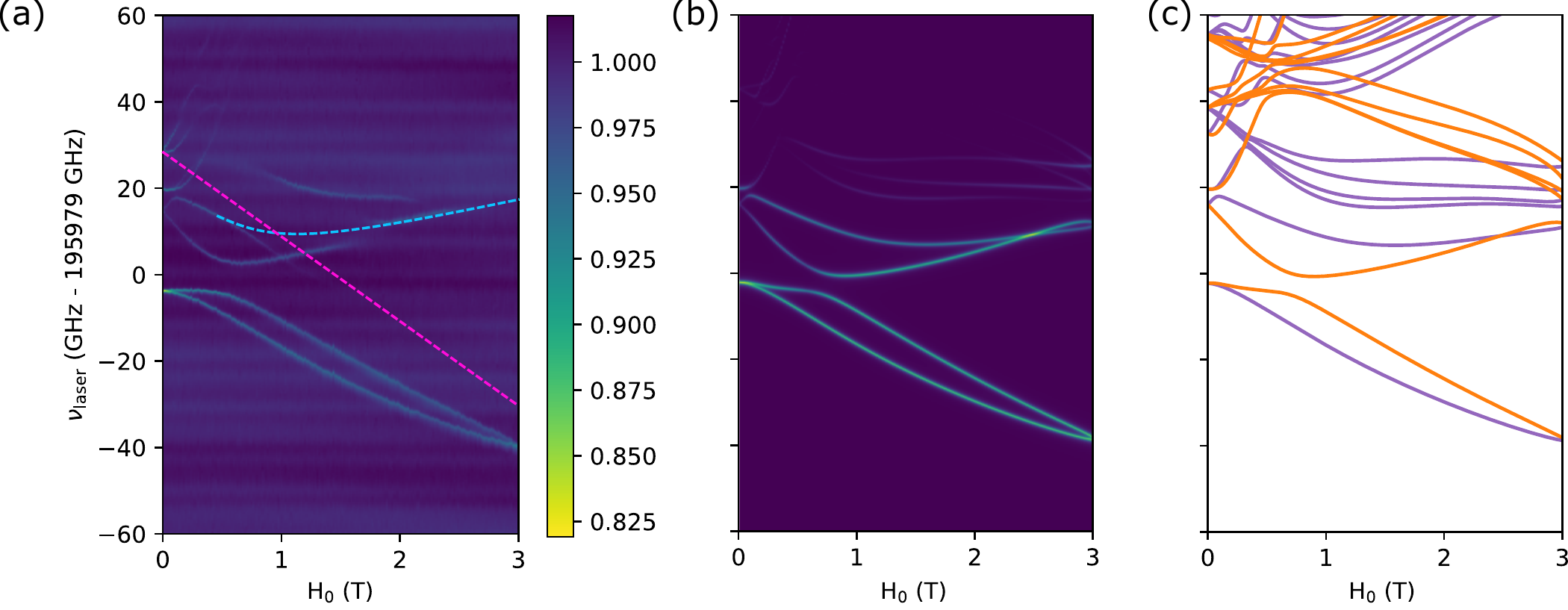}
    \caption{(a) Optical transmission spectrum with light polarised halfway between the \(a\) and \(c\) axes of the crystal and applied magnetic field misaligned significantly from the \(c\)-axis. The purple dashed line is a proposed energy level from something absent from our model, that undergoes and avoided crossing with the erbium level predicted by our model (approximately the blue dashed line) to give rise to the observed spectrum. (b) Output of our cluster model, with relative transition strengths estimated using equation \ref{eq:strengths}, with field  misalignment set to \qty{60}{\degree} from the \(c\)-axis. (c) Model output without transition strengths being taken into account. Orange lines indicate that the erbium is surrounded by gadolinium ions on the `down' sublattice, while purple lines indicate that the erbium is surrounded by gadolinium ions on the `up' sublattice. Here `up' and `down' point along the crystallographic \(c\)-axis.}
    \label{fig:Off-axis}
\end{figure*}

In Figure \ref{fig:Off-axis}(a) we show an optical transmission spectrum of Er:GdVO\(_4\) with a larger misalignment between the crystal and the applied magnetic field. This misalignment was crudely measured to be \qty[separate-uncertainty]{35+-5}{\degree}, from one of the principal axes (either \(a\) or \(c\)). In Figure \ref{fig:Off-axis}(b) we show the output of our model, for external field misalignment chosen to be \(\theta=\qty{60}{\degree}\). This misalignment angle was chosen because it best matched the separation of the two `legs' of the lowest transition doublet, as well as the coupling between the second and third doublets. The model fails to capture the self-crossing of the second doublet at 1 T that is apparent in the data, however, it does manage the higher frequency behaviour around the third doublet. As with the small misalignment case, the fourth transition is predicted to be much weaker than what we observe, but again we see that it is still present in the model (see Figure \ref{fig:Off-axis}(c)).

\section{Discussion}\label{sec:Discussion}
Comparison of the fitted parameters we find in Table \ref{tab: Cluster} to those in Table S1 of Ref. \cite{Hiraishi_2025} shows that our new fitted crystal field parameters only vary by ~\(2\) cm\(^{-1}\) at most. Furthermore, by treating the \(J_\text{eff}=-0.38\) cm\(^{-1}\) found in our previous work as the exchange coupling between the erbium ion and two spin-\(\frac{1}{2}\) Gd\(^{3+}\) ions, then we can estimate the equivalent coupling between the erbium ion and one spin-\(\frac{7}{2}\) Gd\(^{3+}\) ion to be:
\begin{equation}
\begin{split}
    J_\text{eff}^{7/2} & = \frac{2}{4}\cdot\frac{1/2}{7/2}\cdot J_\text{eff} \\
    & = -0.027 \ \text{cm}^{-1}
\end{split}
\end{equation}
This is very close to the fitted value of \(J = -0.0252\) cm\(^{-1}\) that we find for the cluster model. Accordingly, we see that the two models are consistent with each other, with the cluster model having a wider range of validity that extends into regions of high field and to cases where the magnetic field is misaligned with the crystal \(c\)-axis by a significant amount.
The cluster model accomplishes this extended validity with \emph{fewer} parameters. The empirical parameters of Ref.~\cite{Hiraishi_2025} encompassing the magnetic environment $J_\mathrm{eff}$, $B_\mathrm{Gd}$, $B_\mathrm{ex}$, $B_\mathrm{dip}$ are replaced by parameters with clearer physical origins: $J$, $I(H_\text{E})$, $D(H_\text{A})$. 

The fact that our model predicts the fourth transition to be very weak, despite it appearing quite strongly in the measured spectrum, suggests that this transition is not caused by magnetic dipole absorption of the light. 
The four strongest transitions at zero field are from the ground state to the four lowest energy levels of the erbium $^4I_{13/2}$ manifold. Each of these levels has a different crystal-field quantum number $\mu$. The selection rules for these transitions in $\text{D}_{2d}$ symmetry are that one transition is only magnetic-dipole allowed, two are magnetic- and electric-dipole allowed, and the final transition is only electric-dipole allowed. 

The only significant discrepancy between our model and the experimental data is that it misses a minor feature in the off-axis behaviour of the second transition as the magnetic field increases. The data (Figure \ref{fig:Off-axis}(a)) shows the two branches of the second transition crossing at \(\sim  1\) T, while our model does not predict them crossing until \(\sim 2.5\) T. One explanation for this discrepancy is that the measured spectrum would match the model, except the second level is being influenced by an avoided crossing with something that is absent from our model (see the dashed lines in Figure \ref{fig:Off-axis}(a)). This is supported by the fact that the model matches with the data at low and high fields, and only in intermediate values do the two spectra differ.

From our fit we note that \(I\approx2J\), indicating a weaker coupling between the erbium and gadolinium ions. We propose the following explanation for this. Er\(^{3+}\) has a larger nuclear charge than Gd\(^{3+}\), resulting in its \(4f\) electrons being drawn closer to the nucleus. This results in less overlap of the \(4f\) electrons with the orbitals on the adjacent oxide ions, thereby reducing the size of the superexchange coupling \cite{Rado_and_Suhl_Ch2} between the erbium ion and its nearest neighbour gadolinium ions.

Ultimately, the model we present in this paper represents a significant advancement in our understanding of the coupled erbium-gadolinium lattice system. This will be a useful tool for predicting avoided crossings between gadolinium and erbium excitations in the microwave region of the spectrum, which would be good candidates for the observation of microwave to optical transduction.

\section{Acknowledgements}

This work was supported by Quantum Technologies Aotearoa, a research programme of Te Whai Ao – the Dodd-Walls Centre, which is funded by the New Zealand Ministry of Business, Innovation and Employment through International Science Partnerships (Contract No. UOO2347).

\section{Data Availability Statement}

The data that support the findings of this study are available from the corresponding author upon reasonable request.

\section{Code Availability Statement}

The crystal field modelling was carried out using the \texttt{dieke} \cite{dieke} package in Python. Code that reproduces Figures \ref{fig:On-axis} and \ref{fig:Off-axis} is available from the corresponding author upon reasonable request.

\renewcommand{\selectlanguage}[1]{}

\bibliography{reference}

\end{document}